\documentclass[epj]{svjour}
\usepackage{graphicx}
\usepackage{amssymb}
\begin{document}
%\linenumbers
%
\titlerunning{Study of azimuthal Correlations ...}
\title{Study of Azimuthal Correlations in the Target Fragmentation Region in p, d, He, C +
C, Ta and C+Ne, Cu Collisions at a Momentum of 4.2, 4.5 and 10 AGeV/c}
%\string\titlerunning{Study of collective flows...}

\author{L. Chkhaidze\inst{1}
\thanks{\emph{corresponding author E-mail:} ichkhaidze@yahoo.com}
\and  G. Chlachidze\inst{2} \and
T. Djobava\inst{1} \and  A. Galoyan\inst{3} \and  L. Kharkhelauri\inst{1}
\and  R. Togoo\inst{4} \and  V. Uzhinsky\inst{5}
\thanks{\emph{corresponding author E-mail:} uzhinsky@jinr.ru}
% \thanks{\emph{Present address:} ichkhaidze@yahoo.com}
}                     % Do not remove
%
%\offprints{} % Insert a name or remove this line
%
\institute{High Energy Physics Institute, I. Javakhishvili Tbilisi State University, Georgia
 \and Fermi National Accelerator Laboratory, Batavia, Illinois 60510, USA
 \and Veksler and Baldin Laboratory of High Energy Physics, Joint Institute for Nuclear Research,
      Dubna, Russia
 \and Institute of Physics and Technology of the Mongolian Acad. Sci., Ulan Bator, Mongolia
 \and Laboratory of Information Technologies, Joint Institute for Nuclear Research, Dubna, Russia }
\date{Received: date / Revised version: date}
\abstract{Azimuthal correlations between the same type of particles (protons or pions) in the target
fragmentation region was studied in d, He, C + C, Ta (4.2 AGeV/c), C + Ne, Cu (4.5AGeV/c) and p + C, Ta
(10 GeV/c) interactions. The data were obtained from the SKM-200-GIBS streamer chamber and from Propane
Bubble Chamber (PBL-500) systems utilized at JINR. Study of multiparticle azimuthal correlations offers
unique information about space-time evolution of the interactions. Azimuthal correlations were investigated
by using correlation function C($\Delta\phi$)=dN/d($\Delta\phi$), where $\Delta\phi$ represents the angle
between the sums of transverse momenta vectors for particles emitted in the forward and backward hemispheres.
For protons a "back-to back" ("negative") azimuthal correlations were observed in the above mentioned
interactions. The absolute values of the correlation coefficient $|\xi|$ -- the slope parameter of
C($\Delta\phi$), strongly depend on the mass number of the target ($A_T$) nuclei in the nucleon-nucleus
and nucleus-nucleus collisions. Namely, $|\xi|$ -- decreases with increase of $A_T$ in p+C and p+Ta collisions,
while $|\xi|$ decreases from d+C up to C+Ne and then almost does not change with increase of $A_P$, $A_T$ in
(d+He)Ta, C+Cu and C+Ta collisions. For pions a "back-to-back" correlations were obtained for a light targets
(C, Ne), and a "side-by-side" ("positive") correlations for a heavy targets (Cu, Ta). The $|\xi|$
insignificantly changes with increase of the momenta per nucleon and almost does not change with increase
of $A_P$ and $A_T$. Models, used for description of the data -- the Ultra relativistic Quantum Molecular
Dynamic (UrQMD) and Quark-Gluon String Model (QGSM), satisfactorily describe the obtained experimental results.
\PACS{
      {25.70.-z}{}
\and
      {25.75.Ld}{}
     % end of PACS codes
 }
} %end of abstract
\maketitle
\section{Introduction}
Studies of multi-particle correlations have provided crucial insight into the underlying mechanism of
particle production in relativistic heavy-ion collisions. The primary goal of current relativistic heavy
ion research is the creation and study of nuclear matter at high energy densities \cite{R1,R2}. Open
questions  the detailed properties of such excited matter, as well as a transition to the quark-gluon
plasma (QGP) phase. Such a phase of deconfined quarks and gluons has been predicted to survive for $\sim $
3--10 fm/c in Au-Au collisions at the Relativistic Heavy Ion Collider (RHIC) \cite{R3} and several experimental
probes have been proposed for its possible detection and study \cite{R4w}. The most prominent feature of
multi-particle correlations in AA collisions is due to collective flow \cite{R4,R5}, an azimuthal anisotropy
in momentum space induced by strong expansion of the initial almond-shaped overlap area of two nuclei \cite{R4}.
Collective flow constitutes an important observable \cite{R6} because it is thought to be driven by pressure
built up early in the collision, and therefore can reflect conditions exciting in the first few fm/c.
Collective flow leads to an anisotropy in the azimuthal distribution of emitted particles. Studies of elliptic
flow have been carried out over a wide range of energies and systems at both RHIC and the LHC \cite{R2,R7}--\cite{R10w}.
Study of multiparticle azimuthal correlations offers unique information about space-time evolution of
the collective systems \cite{R11,R12}. One of the interesting methods proposed in \cite{R13,R14} is the
conventional division of phase space into forward and backward moving particles according to the rapidity
and emission angle or into slow and fast particles according to energy.

During last years we have analyzed our experimental data following the method \cite{R13,R14}. The collective
variables and their dependencies on the transverse momentum of all secondary charged particles in the azimuthal
plane, were studied to reveal a nontrivial effects in nucleus-nucleus collisions \cite{R15}--\cite{R19}. We have
investigated multiparticle azimuthal correlations of protons and pions in central and minimal bias inelastic
collisions (4.2, 4.5 and 10 GeV/c/nucleon) within two experimental setups -- 2 m streamer chamber placed in a
magnetic field (SKM-200-GIBS) and 2 m Propane Bubble Chamber (PBC-500) of JINR. In order to investigate
the mechanism of nucleus-nucleus interactions we have studied the correlations of these particles (protons,
pions, $\Lambda$-hyperons) with respect to the reaction plane (directed and elliptic flows) \cite{R15}--\cite{R17}, as
well as with respect to the opening angle between particles emitted  in the forward and backward hemispheres
\cite{R18,R19}.

In this paper, we present results of the analysis of azimuthal multiparticle correlations in the target
fragmentation region in d, He, C + C, Ta (4.2 AGeV/c), C + Ne, Cu (4.5AGeV/c) and p + C, Ta (10 GeV/c)
collisions between the same type of particles (protons or pions). The dependence of the azimuthal
correlation coefficient on the mass numbers of projectile ($A_P$) and target ($A_T$)  have been investigated.

\section{ Experimental data}
The data were obtained from the SKM-200-GIBS streamer chamber and from the Propane Bubble Chamber systems
(PBC-500) utilized at JINR. The SKM-200-GIBS setup is based on a 2 m streamer chamber placed in the magnetic
field of 0.8 T and on a triggering system. C+Ne and C+Cu central collisions were studied only at the
momentum of 4.5 AGeV/c, because the multiplicity of the second particles significantly increases using
a central trigger. A central trigger selected events with no charged projectile spectator fragments at p $>$ 3
GeV/c, within a cone of half angle of either $\Theta_{ch}$ = 2.4$^\circ$ or 2.9$^\circ$, depending on
the run. The thickness of Cu solid target, in the form of a thin disc, was 0.2 g/cm$^2$. Neon gas filling
of the chamber \cite{R20} also served as a nuclear target. Details of the acquisition techniques
of 723 C+Ne and 663 C+Cu interaction data, and other experimental procedures have been presented in
\cite{R20,R21}.

The 2 meter Propane Bubble Chamber (PBC-500) was placed in the magnetic field of 1.5 T. The procedures for
separating out the p+C, d+C, He+C and C+C collisions in propane (${\rm C}_3{\rm H}_8$) and the processing
of the data including particle identification and corrections have been described in detail in Refs.
\cite{R22}--\cite{R24}. The analysis produced 16509 p+C, 4581 d+C, 9737 He+C, 15962 C+C,  2342 p+Ta, 1424 d+Ta,
1532 He+Ta (at that d+Ta and He+Ta data were combined to increase statistics) and 2469 C+Ta inelastic
collision events. The protons with momentum p$<$150 MeV/c were not detected within the PBC-500 (as far as
their track lengths $l\ <$ 2 mm) and protons with p$\ <$ 200 MeV/c were absorbed in Ta target plate
(the detector biases).

In the experiment, the projectile fragmentation products were identified as those characterized by the momentum
p $>$ 3.5 GeV/c (4.2, 4.5 A GeV/c) or p $>$ 7 GeV/c (10 GeV/c) and polar angle $\Theta\ <$ 3.5$^\circ$ and
the target fragmentation products -- by the momentum p $<$ 0.25 GeV/c in the target rest frame. After these
selection criterions, the remaining protons were considered as the participant protons. For the analysis
minimum three particles $N_{particles} \geq$ 3 in the event were required.

\section{Azimuthal Correlations Between Protons or Pions}
In Refs. \cite{R13,R14} the method for studying the correlation between groups of particles has been developed.
The azimuthal correlation function was defined \cite{R13,R14} by the relative opening angle between
the transverse momentum vector sums of particles emitted forward and backward with respect to the rest frame
of the target nucleus (larger or smaller a rapidity of $y_0$=0.2). The method was used for analysis of the data
at 4.9, 60 and 200 GeV (BEVALAC, CERN/SPS) \cite{R14}.

We apply this method for our data. The analysis has been performed event by event. In each event, we denote
the vectors:
\begin{equation}
\vec Q_B\ =\ \sum_{y_i<y_0}\vec P_\bot \label{Eq1}
\end{equation}
and
\begin{equation}
\vec Q_F\ =\ \sum_{y_i\geq y_0}\vec P_\bot, \label{Eq2}
\end{equation}
where $y_0$ is 0.2 and $y_i\ \leq$ 1.33 for protons. This is guided by the experimental fact that the target
rapidity distributions of protons peaks at  $y_0 \simeq 0.2$ \cite{R25} and follows the idea of a target
"fireball" moving with a rapidity $y_0$ (in the rest frame of the target nucleus). The influence of the actually
chosen value of $y_0$ to the experimental results, furthermore, has been checked by varying $y_0$ in a reasonable
range $0.1\leq y_0\leq 0.3$ and within these limits no significant change of the correlation function has been
observed. The correlation function C($\Delta \phi$) was determined as:

\begin{equation}
C(\Delta \phi) = dN/d\Delta \phi \label{Eq3},
\end{equation}
where $\Delta \phi$ is the angle between the vectors $\vec Q_B$ and $\vec Q_F$:
\begin{equation}
\Delta \phi = \arccos \frac{(\vec Q_B \cdot \vec Q_F)}{(|\vec Q_B| \cdot |\vec Q_F|)}. \label{Eq4}
\end{equation}

Essentially, C($\Delta \phi$) measures whether the particles are preferentially emitted "back-to back" or
"side-by-side" correlations. "Back-to back" means the "negative" correlations, where C($\Delta \phi$) increases
with $\Delta \phi$ and reaches a maximum at $\Delta \phi$ = 180$^\circ$.  "Side-by-side " means the "positive"
correlations, where C($\Delta \phi$) decreases with $\Delta \phi$ and have a maximum at $\Delta \phi$ = 0$^\circ$
\cite{R13,R14}.

In view of the strong coupling between the nucleons and pions, it is interesting to know the correlations
between pions. Thus, we have studied correlations between protons and between pions. Fig. 1--3 show the
experimental correlation function C($\Delta \phi$) for these particles in p, d, He, C + C, p, d, He, C + Ta
and C + Ne, Cu collisions. One can observe from Figures a clear correlation for protons and for pions
(The results obtained for better performance, specifically, the corresponding C($\Delta \phi$) -distributions
of protons are shifted below for C + Ne, Cu, Ta collisions by 0.85 factor). To quantify these experimental
results, the data were fitted by the function:
\begin{equation}
 C(\Delta \phi) = a\ (1 + \xi cos(\Delta \phi)), \label{Eq5}
\end{equation}
where $a$ is a normalization constant.

Results of the fitting are listed in Table 1. The strength of the correlation is defined as:
\begin{equation}
\zeta = C(0^\circ) / C(180^\circ) = (1 + \xi )/(1 - \xi ). \label{Eq6}
\end{equation}

As it can be seen, the correlation coefficient $\xi<0$, and thus the strength of correlation $\zeta<1$ for
protons in all interactions. This means that protons are preferentially emitted back-to-back.

One can observe from Figs. 1 and 2, a clear back-to-back ($\xi<0$, $\zeta<1$) correlations for pions in light
systems p, d, He, C + C and  C + Ne. For heavy, asymmetric pairs of p, (d+He), C + Ta  and C + Cu the
side-by-side ($\xi>0$ and $\zeta>1$) correlations of pions can be seen from Fig. 3 (Table 1).

We have studied also a dependence of the correlation coefficient $|\xi|$ on  $A_P$, $A_T$ for protons and pions.
The absolute values of $\xi$ for protons decreases with increase of $A_T$ in the nucleon-nucleus collisions:
0.302$\pm$0.013 for p+C up to 0.143$\pm$0.027 for p+Ta (10 GeV/c). What concerns nucleus-nucleus collisions,
$|\xi|$ initially linearly decreases with increase of $A_P$, $A_T$ from 0.296$\pm$0.022 (d+C) up to
0.150$\pm$0.030 (C+Ne), and then almost does not change with increase of $A_P$ , $A_T$. For pions, $|\xi|$
insignificantly changes with increase of the momenta per nucleon and almost does not change with increase of
$A_P$ , $A_T$: $|\xi|$ = 0.107$\pm$0.019 for p+C and it's almost the same for C+Ta (Table 1).

The obtained experimental results for p, d, He, C + C, Ta collisions have been compared to the predictions
of the Ultra relativistic Quantum Molecular Dynamic (UrQMD) model \cite{R26,R27} coupled with the Statistical
Multifragmentation (SMM) model \cite{R28}, and the Quark-Gluon String Model (QGSM) \cite{R29,R30} for C + C,
Ne, Cu, Ta interactions.

The UrQMD model is a microscopic transport model based on the covariant propagation of all hadrons on classical
trajectories in a combination with stochastic binary scatterings, colour string formation and resonance decay.
The SMM model \cite{R28} was added to the UrQMD model \cite{R31,R32} to improve simulations of a baryon
production in target and projectile fragmentation regions. There is not a strong subdivision between the
participating protons and protons created at the de-excitation of nuclear residuals. The SMM allows a simulation
of the evaporated proton production, and UrQMD generates the participating protons. Thus, we have in the model
events of all processes presented in the experiment. The UrQMD model is designed as a multi-purpose tool for
studying a wide variety of heavy-ion-related effects ranging from multi-fragmentation and collective flow to
particle productions and correlations in the energy range from SIS to RHIC.

The experimental selection criteria of events were applied to the events generated by the enlarged UrQMD
model (UrQMD+SMM). 51191 (10 GeV/c) p+C, 27502 (4.2 A GeV/c)  d+C, 31716 (4.2 A GeV/c) He+C, 35754 (4.2 A GeV/c)
C+C, 7230 (10 GeV/c) p+Ta, 5137 (4.2 A GeV/c) d+Ta, 3210 (4.2 A GeV/c)  He+Ta and 9130 (4.2 A GeV/c) C+Ta
interactions have been chosen after the experimental events selection criteria. The values of the correlation
coefficient $\xi$ were extracted for protons and pions from corresponding distributions for each nuclear pair
(Table 1). One can see, that there is a good agreement between the experimental and theoretical values
(Figs. 1--4).

Detailed description of the QGSM can be found in \cite{R29,R30}. This model is based on the Regge and string
phenomenology of particle production in inelastic binary had\-ron collisions. Within QGSM, nuclear densities are
used for selecting coordinates of original nucleons. This is followed by the formation of quark-gluon strings
which fragment into hadrons. Those hadrons rescatter. In QGSM, the sole cause of sidewards flow is the hadron
rescattering. In all other models in the literature, where produced pressure comes from rescattering alone,
the flow in the specific energy regime was found significantly underestimated, albeit predominantly in heavier
systems \cite{R33}. However, in collisions studied here, QGSM turned out to describe the azimuthal correlations
rather well (Figs. 2--4).

For simulating the model events, we have employed the COLLI Monte-Carlo generator \cite{R34} based on QGSM.
To the generated events, the trigger filter has been applied in the case of C + C, Ne, Cu, Ta interactions.
In mimicking, in particular, the deterioration of experimental efficiency for registering vertical tracks,
protons characterized by deep angles greater than 60$^\circ$ have been excluded from an analysis. For the
present study, 36495 C+C, 16637 C+Ne, 8347 C+Cu and 9721 C+Ta collisions at a momentum of 4.2, 4.5 AGeV/c have
been chosen to the analysis (Table 1). The mass dependence, as seen in Fig. 4 is, in particular, fairly well
reproduced. C + C, Ne, Cu, Ta events have been generated at fixed $b$-values equal to: $<b>$  = 2.65, 2.20,
2.75 and 6.54 (fm), respectively.

Finally, both models describe experimental results. As seen, there is a good agreement between the experimental
and MC data. The experimental results from C+C and C+Ta collisions have been compared to the results obtained by
the UrQMD and QGSM models in order to study the accuracy of the mechanism of nucleus-nucleus interactions
implemented in these models. The values of the correlation coefficients in the experimental and MC data for
protons and pions are in a good agreement for all studied systems.

Back-to-back correlations have been observed between protons with the Plastic-Ball detector in p + Au collisions
at energy of 4.9, 60, 200 GeV/c and in (O, S)Au reactions at 200 GeV/c \cite{R14,R35}. Because, the azimuthal
correlation function was defined in the target fragmentation region, the correlation parameters in the wide range
of energy increases inappreciable. We applied this method also in our previous articles \cite{R18,R19}, where the
analysis was carried out in the whole rapidity region instead of the target fragmentation region
\cite{R13,R14}. In the case, instead of $y_0$ -- an average rapidity of participant protons for each colliding pears,
$y_c$ was used.

The back-to back emission of protons can be understood as results of (local) transverse momentum conservation
\cite{R14}.
Back-to-back ($\xi < 0$, $\zeta < 1$) pion correlation for light systems p, d, He, C + C and C + Ne, and  the
side-by-side pion azimuthal correlations ( $\xi > 0$ and $\zeta > 1$) for heavy, asymmetric pairs -- p, d, He,
C + Ta and C + Cu, i. e. change of the pion's azimuthal correlation type, are in agreement with our previous
results \cite{R18,R19}. Also, the side-by-side correlations of pions have been observed in p+Au collisions at
Bevalac (4.9 GeV/nucleon) and CERN-SPS (60 and 200 GeV/nucleon) energies \cite{R14}. Another investigations of
large angle two-particle correlations, carried out at the 3.6 AGeV C-beam in Dubna \cite{R12}, showed a
back-to-back pion correlation for a light target (Al), and a side-by-side correlation for a heavy target (Pb).
For protons a back-to-back correlation was observed for all targets. All above mentioned results appear to be
consistent with our results of proton and pion azimuthal correlations and with the variation of the pion
correlation when going from C to Ta targets.

The reason for the observed difference between protons and pions is that the pions are absorbed in the excited
target matter ($\pi + N\rightarrow \Delta$ and $\Delta + N \rightarrow N + N$) \cite{R13,R14}. The side-by-side
correlation of pions can be naturally explained on the base that pions, which are created in collision suffer at
$b\ \neq\ 0$ fm ($b$ is the impact parameter) either rescattering or even complete absorption in the target
spectator matter. Both processes will result in a relative depletion of pions in the geometrical direction of
the target spectator matter and hence will cause an azimuthal side-by-side correlation as observed in
the experimental data.

\section{Conclusion}
The study of multiparticle azimuthal correlations between protons or between pions in the target fragmentation
region in d, He, C + C, Ta (4.2 A GeV/c), C + Ne, Cu (4.5 A GeV/c) and p + C, Ta (10 GeV/c) collisions have been
carried out. The pC system is the lightest studied one, and the pTa is extremely asymmetrical system in which
azimuthal correlations between protons and between pions in the "backward" and "forward" hemispheres have
detected:
\begin{enumerate}
\item For protons a bach-to-back correlation was observed for all interactions. The absolute values of $\xi$
for protons decreases with increase of the mass numbers of target $A_T$ nuclei in the nucleon-nucleus collisions:
0.302$\pm$0.013 for p+C  up to 0.143$\pm$0.027 for p+Ta at 10GeV/c. What concerns nucleus-nucleus collisions,
$|\xi|$ initially linearly decreases with increase of the mass number of projectile $A_P$ and target $A_T$ nuclei
from 0.296$\pm$0.022 (d+C) up to 0.150$\pm$0.030 (C+Ne) and then almost does not change with increase of $A_P$,
$A_T$.
\item As shown, the pions exhibit "back-to back" (negative) correlations consistent with that for protons
for light systems of p, (d+He), C + C and C + Ne (light targets -- C, Ne) and a side-by-side correlations for
heavy, asymmetric pairs of p, d, He, C + Ta and C+Cu (heavy targets -- Cu, Ta). The correlation coefficient
($|\xi|$) insignificantly changes with increase of the momenta per nucleon and almost does not change with
increase $A_P$, $A_T$ nuclei: $|\xi|$ = 0.107$\pm$0.019 for p+C, and it's almost the same for C+Ta.
\item Both models, UrQMD and QGSM satisfactorily describe the experimental results, there is a good agreement
between the experimental and theoretical values.
\end{enumerate}

\section* {Acknowledgements}
\vspace{.5cm} This work was partially supported by the Georgian Shota Rustaveli National Science Foundation under
Grant DI/38 /6-200/13.

The authors are thankful to heterogeneous computing (HybriLIT) team of the Laboratory of Information
Technologies of JINR  for support of our calculations.  A.G. and V.U. are grateful to A. Botvina for providing
us with the SMM code.

%
% BibTeX users please use
% \bibliographystyle{}
% \bibliography{}

\newpage
\onecolumn
{\bf Figure and Table captions}

\vspace{5mm}
Fig.1. The dependence of the correlation function C($\Delta \phi$) on the $\Delta \phi$ for protons
and pions in experimental ($\bullet$, $\blacktriangle$) and UrQMD ($\circ$, $\bigtriangleup$) (pC, dC, HeC, CC),
QGSM ($\ast$, $\star$) (CC) MC data, respectively. The curves are the results of the approximation of the data
(see text).

\vspace{5mm}
Fig.2. The dependence of the correlation function C($\Delta \phi$) on the $\Delta \phi$ from CNe
collisions for protons and pions in experimental ($\bullet$, $\blacktriangle$) and QGSM ($\ast$, $\star$) MC data,
respectively. The curves are the results of the approximation of the data (see text).

\vspace{5mm}
Fig.3. The dependence of the correlation function C($\Delta \phi$) on the $\Delta \phi$ for protons
and pions in experimental ($\bullet$, $\blacktriangle$) and UrQMD ($\circ$, $\bigtriangleup$) (p, d+He)Ta,
(CTa), QGSM ($\ast$, $\star$) (CCu, CTa) MC data, respectively. The curves are the results of the approximation
of the data (see text).

\vspace{5mm}
Fig.4. The dependence of $|\xi|$ correlation coefficient on ($A_P + A_T)^{1/2}$ for protons (upper points)
and pions (lower points) in pC, dC, HeC,  CC, pTa, CNe, (dHe)Ta, CCu and CTa collisions. Closed symbols
correspond to experimental ($\bullet$, $\blacktriangle$), and open ones to the UrQMD ($\circ$, $\bigtriangleup$),
stars to the QGSM ($\ast$, $\star$) MC data, respectively.

\vspace{5mm}
Table 1. The number of experimental and Monte-Carlo (MC) generated (UrQMD and QGSM) events
($N_{event}$), the correlation coefficient ($\xi$) for protons and pions in these collisions.

\newpage
%%%Table 1%%%%%%%%%%%%%%%
\begin{table}
\caption{}
\begin{center}
\begin{tabular}{|c|c|c|c|c|}
\hline

$A_{P}$, $A_{T}$ & $N_{events}$ & $\xi_{prot}$ & $\xi_{pion}$ &   $\sqrt{A_P*A_T}$            \\ \hline
    pC           & $\begin{array}{lr}exp.& 16509\\ UrQMD& 50191\end{array}$
                 & $\begin{array}{c}-0.302 \pm 0.013 \\ -0.298 \pm 0.012 \end{array}$
                 & $\begin{array}{c}-0.107 \pm 0.019 \\ -0.112 \pm 0.012 \end{array}$ &  3.50 \\ \hline

    dC           & $\begin{array}{lr}exp.& 4581\\ UrQMD& 27502\end{array}$
                 & $\begin{array}{c}-0.296 \pm 0.022 \\ -0.306 \pm 0.009 \end{array}$
                 & $\begin{array}{c}-0.108 \pm 0.028 \\ -0.104 \pm 0.010 \end{array}$ &  4.93 \\ \hline

   HeC           & $\begin{array}{lr}exp.& 9737\\ UrQMD& 31716\end{array}$
                 & $\begin{array}{c}-0.259 \pm 0.021 \\ -0.252 \pm 0.016 \end{array}$
                 & $\begin{array}{c}-0.103 \pm 0.018 \\ -0.109 \pm 0.012 \end{array}$ &  6.93 \\ \hline

    CC           & $\begin{array}{lr}exp.& 15962\\ UrQMD& 35754\\ QGSM& 36495\end{array}$
                 & $\begin{array}{c}-0.163 \pm 0.010 \\ -0.161 \pm 0.014 \\ -0.174 \pm 0.009 \end{array}$
                 & $\begin{array}{c}-0.075 \pm 0.011 \\ -0.076 \pm 0.012 \\ -0.080 \pm 0.010 \end{array}$ &  12.00 \\ \hline

    pTa          & $\begin{array}{lr}exp.& 2342\\ UrQMD& 7230\end{array}$
                 & $\begin{array}{c}-0.143 \pm 0.027 \\ -0.140 \pm 0.014 \end{array}$
                 & $\begin{array}{c} ~~~0.111 \pm 0.021 \\  ~~~0.116 \pm 0.015 \end{array}$ &  13.50 \\ \hline

    CNe          & $\begin{array}{lr}exp.& 723\\ QGSM& 16637\end{array}$
                 & $\begin{array}{c}-0.150 \pm 0.030 \\ -0.147 \pm 0.018 \end{array}$
                 & $\begin{array}{c}-0.081 \pm 0.015 \\ -0.088 \pm 0.014 \end{array}$ &  16.60 \\ \hline

   (d+He)Ta      & $\begin{array}{lr}exp.& 2956\\ UrQMD& 17629\end{array}$
                 & $\begin{array}{c}-0.153 \pm 0.022 \\ -0.148 \pm 0.012 \end{array}$
                 & $\begin{array}{c} ~~~0.086 \pm 0.024 \\  ~~~0.085 \pm 0.014 \end{array}$ &  23.00 \\ \hline

    CCu          & $\begin{array}{lr}exp.& 663\\ QGSM& 8347\end{array}$
                 & $\begin{array}{c}-0.139 \pm 0.035 \\ -0.148 \pm 0.015 \end{array}$
                 & $\begin{array}{c} ~~~0.093 \pm 0.022 \\  ~~~0.082 \pm 0.017 \end{array}$ &  27.70 \\ \hline

    CTa          & $\begin{array}{lr}exp.& 2469\\ UrQMD& 9130\\ QGSM& 9721\end{array}$
                 & $\begin{array}{c}-0.144 \pm 0.025 \\ -0.148 \pm 0.014 \\ -0.160 \pm 0.006 \end{array}$
                 & $\begin{array}{c} ~~~0.108 \pm 0.022 \\ ~~~0.105 \pm 0.016 \\ ~~~0.093 \pm 0.006 \end{array}$ &  46.60 \\ \hline
\end{tabular}
\end{center}
\end{table}
%\end{document}
%------------------------ Fig_1----------------------------
%%%Figure 1
\begin{figure}[bth]
\begin{center}
\includegraphics[width=70mm,height=70mm,clip]{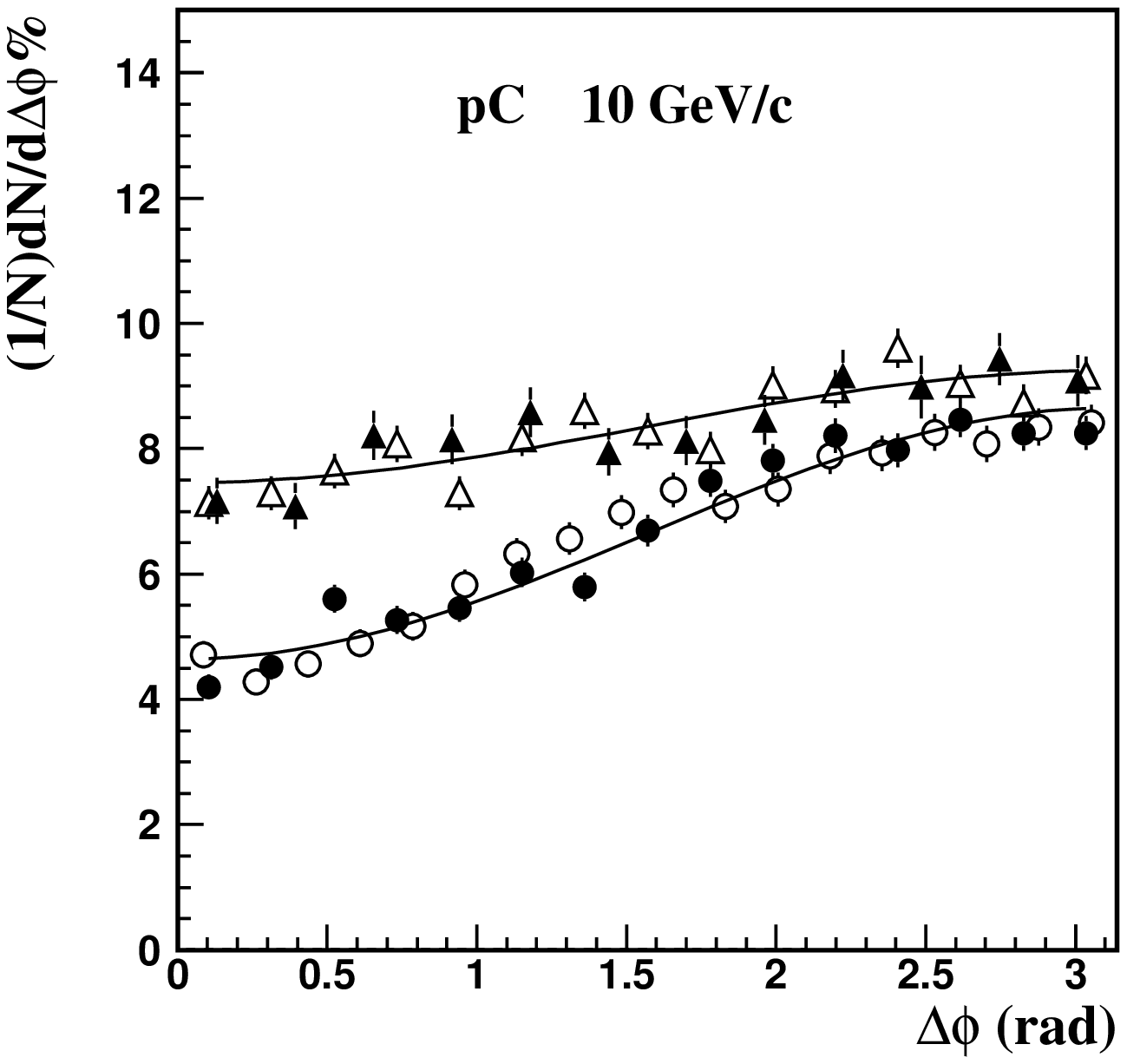}~~\includegraphics[width=70mm,height=70mm,clip]{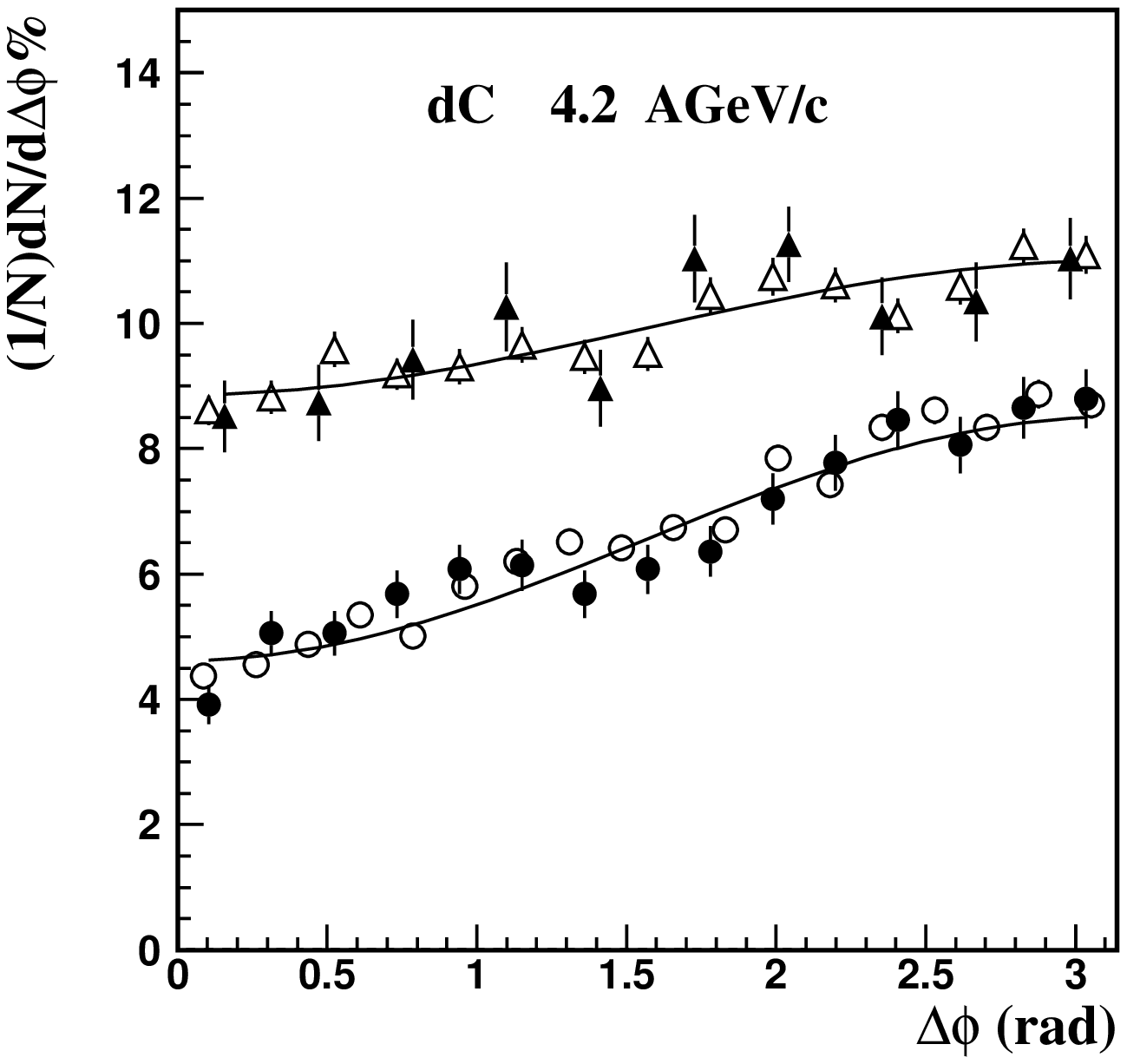}
\includegraphics[width=70mm,height=70mm,clip]{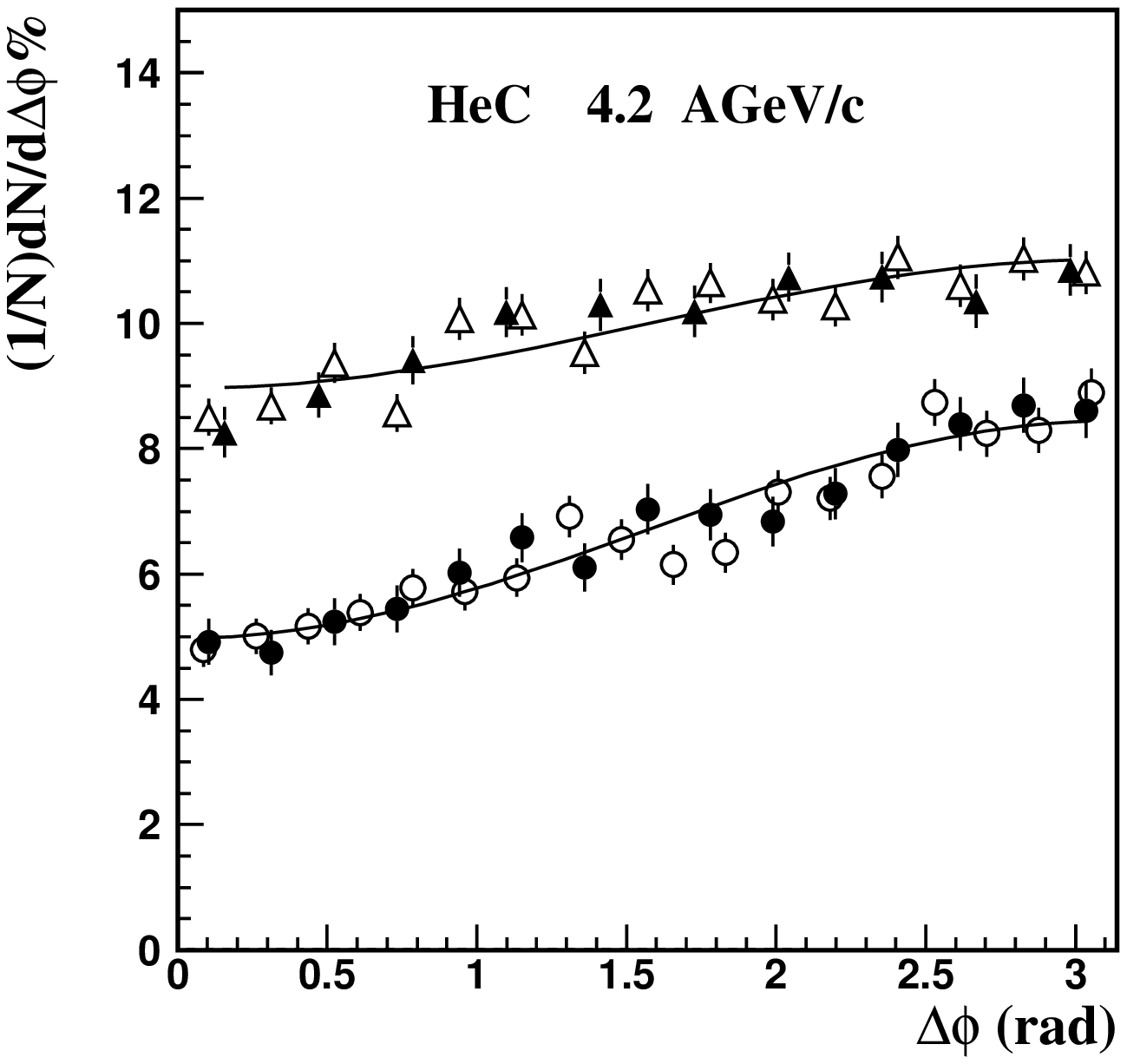}~~\includegraphics[width=70mm,height=70mm,clip]{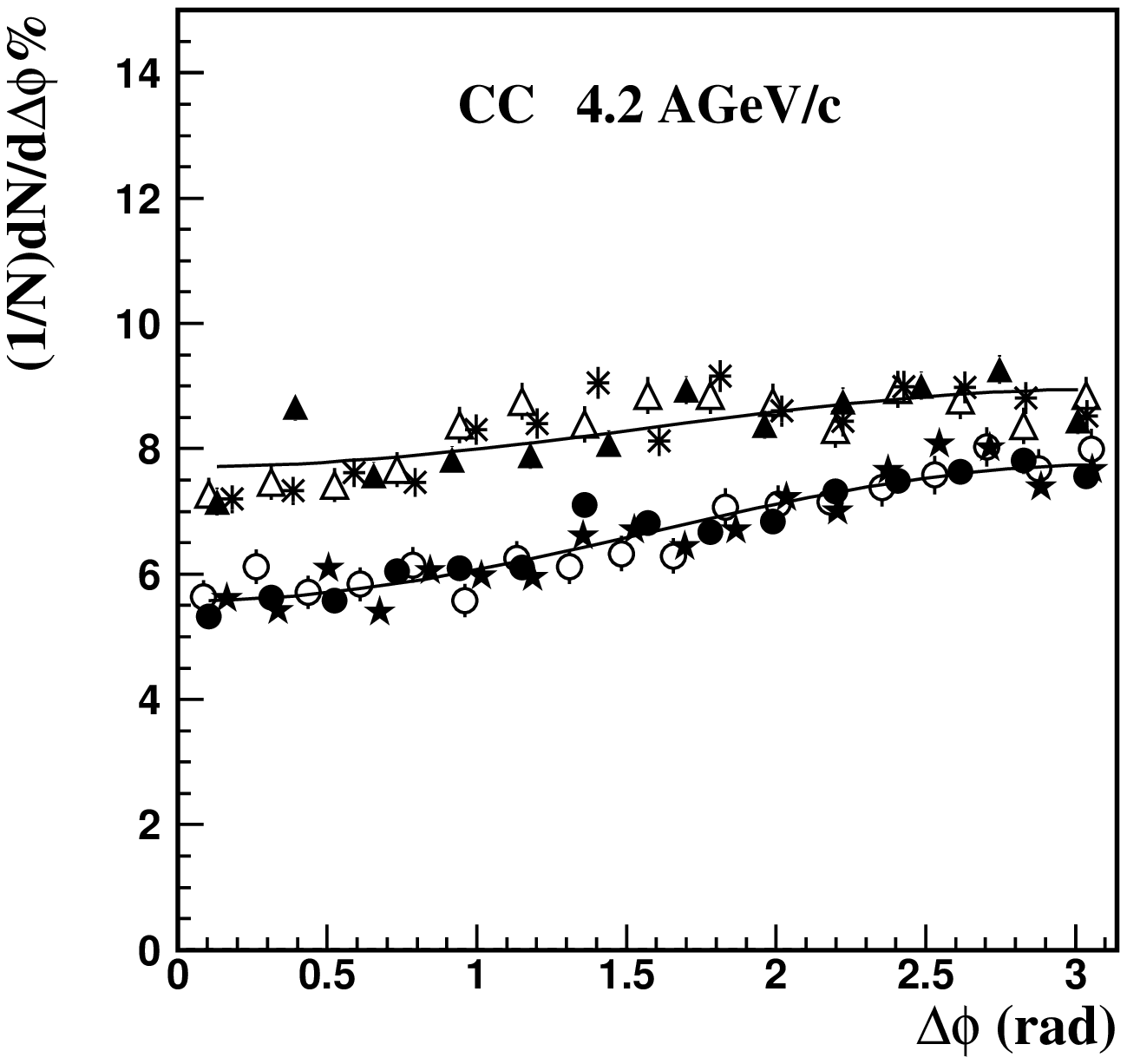}
\caption{}
\end{center}
\end{figure}

%FIGURE 2
\begin{figure}[bth]
\vspace{3cm}
\begin{center}
\includegraphics[width=70mm,height=70mm,clip]{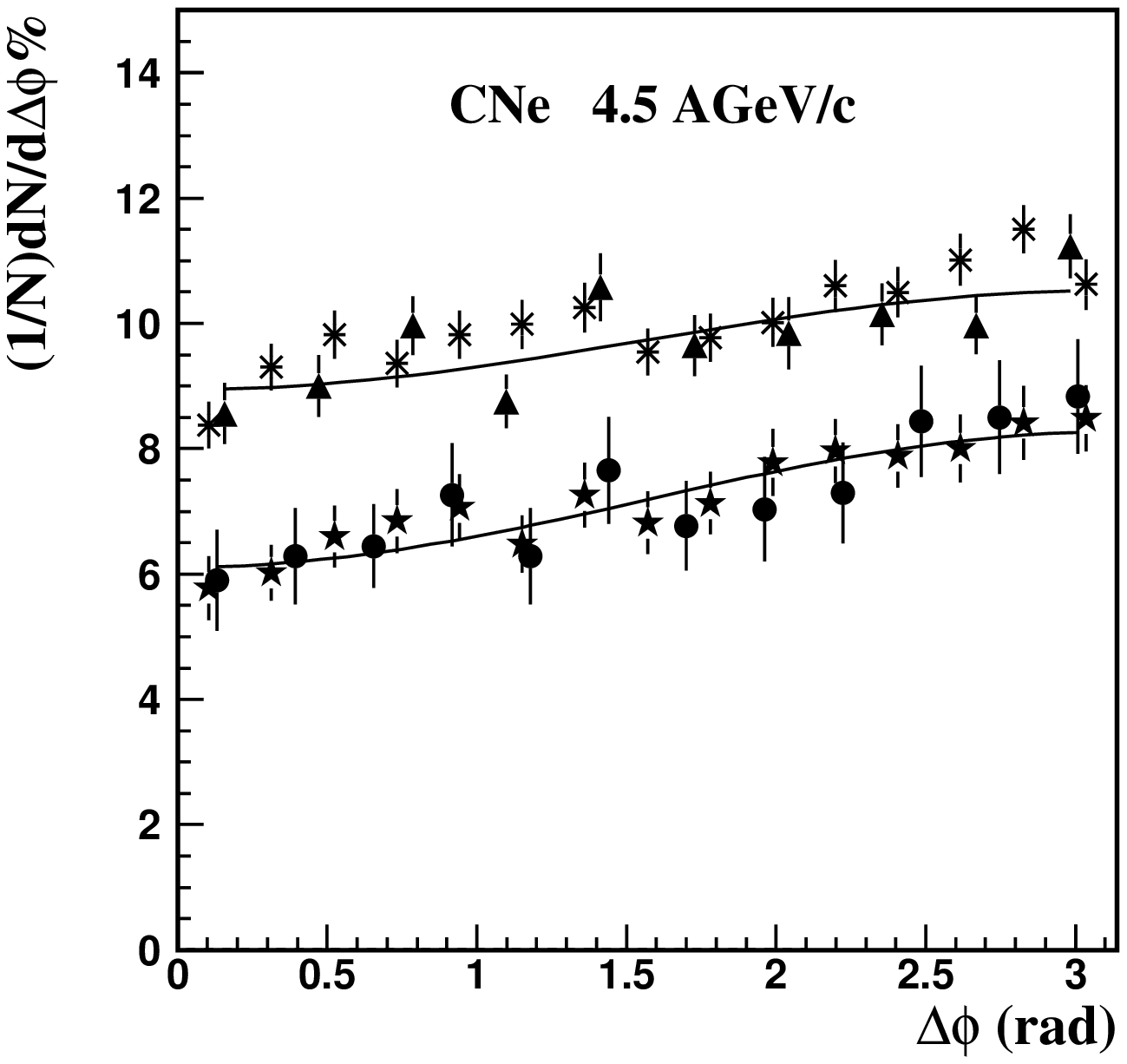}
\caption{}
\end{center}
\end{figure}

\newpage

%FIGURE 3
\begin{figure}[bth]
\begin{center}
\includegraphics[width=70mm,height=70mm,clip]{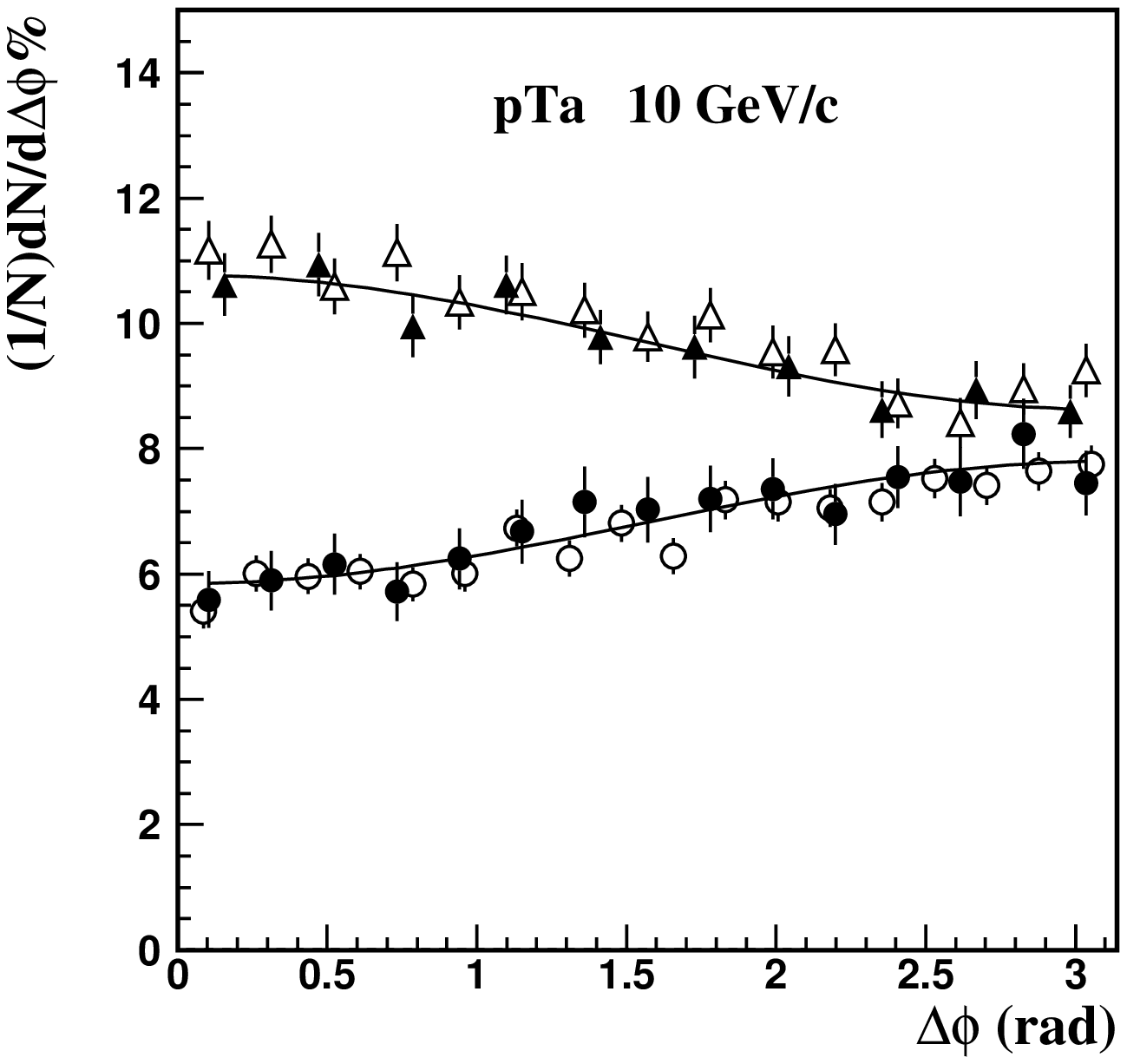}~~\includegraphics[width=70mm,height=70mm,clip]{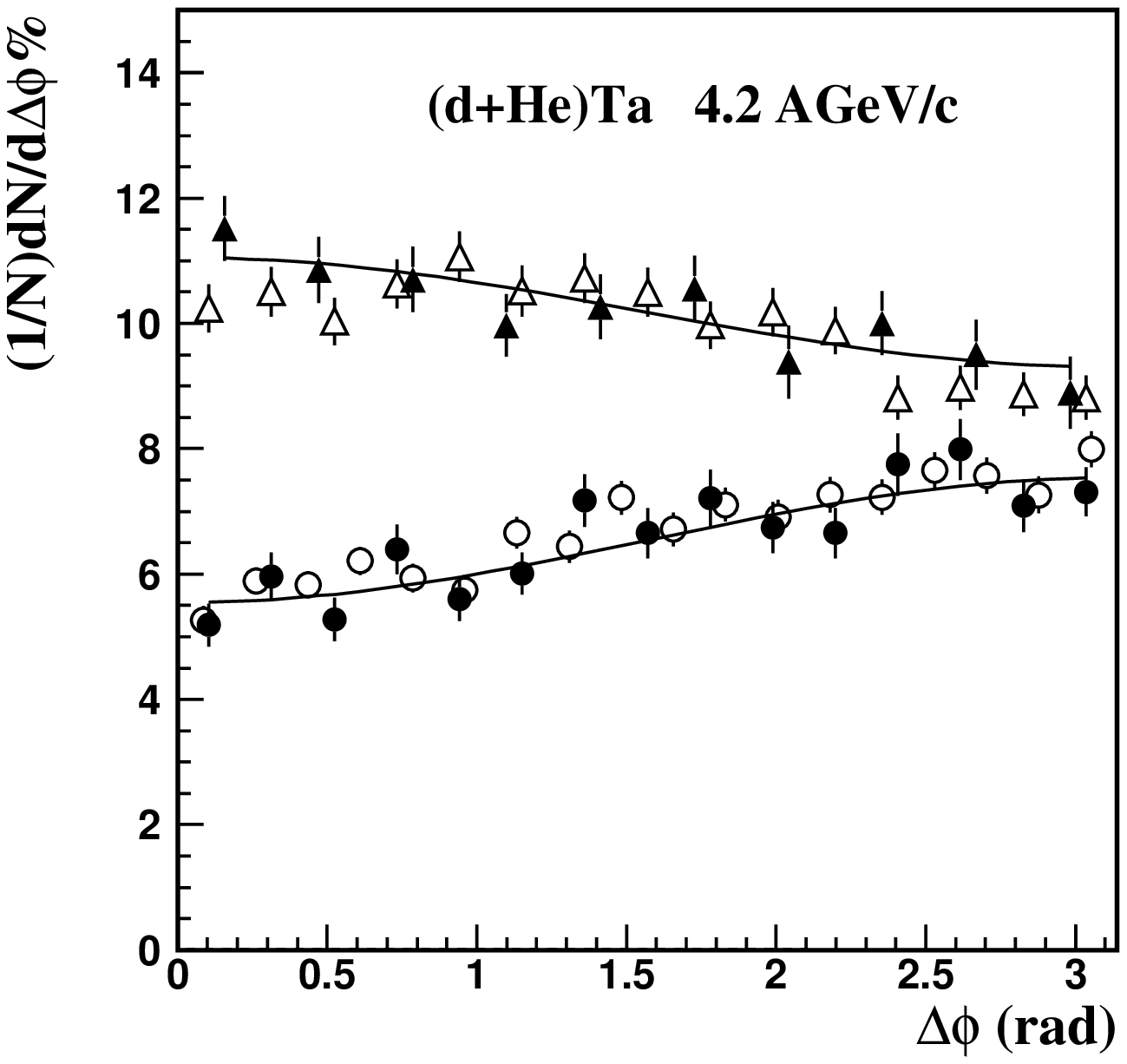}
\includegraphics[width=70mm,height=70mm,clip]{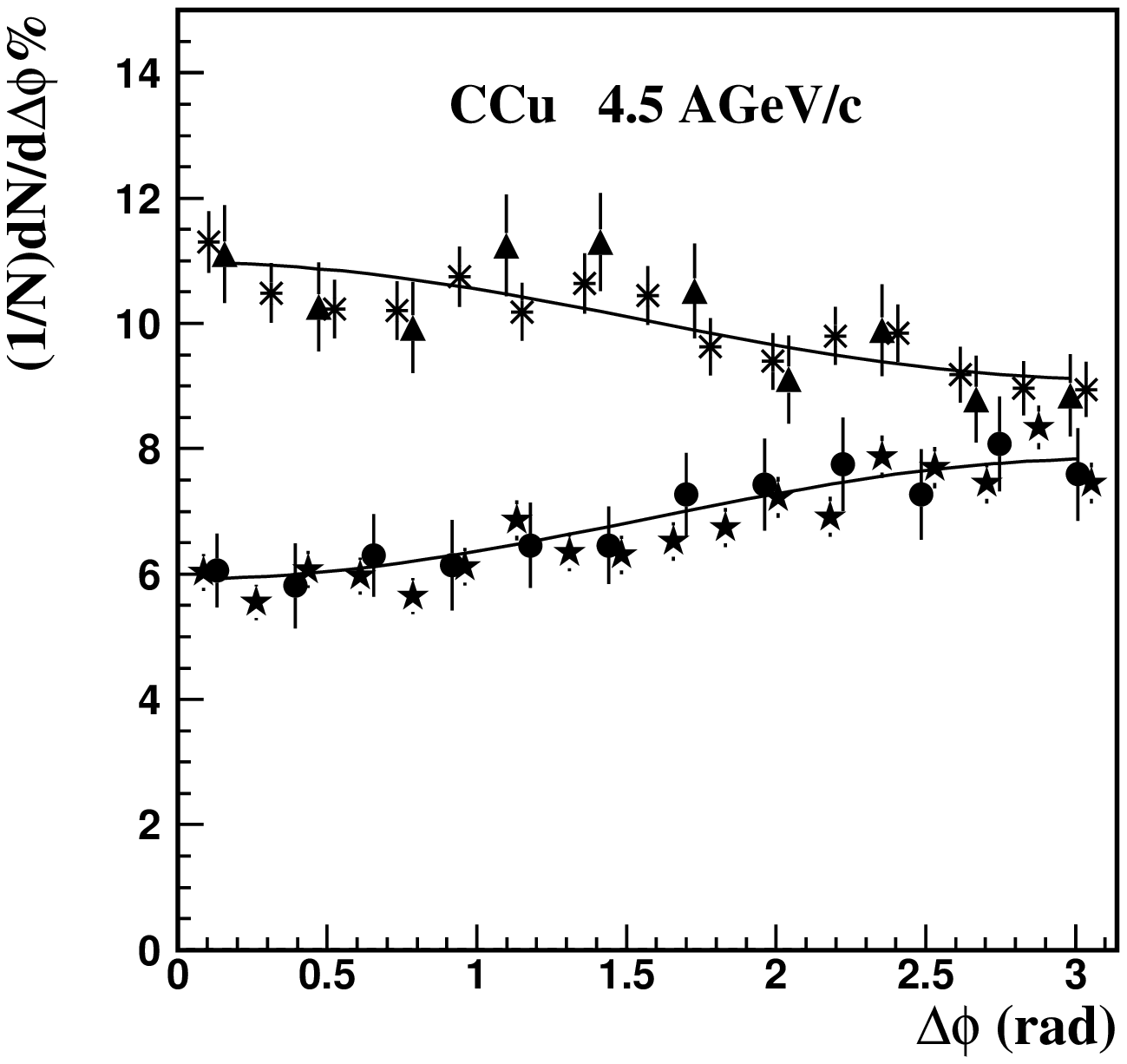}~~\includegraphics[width=70mm,height=70mm,clip]{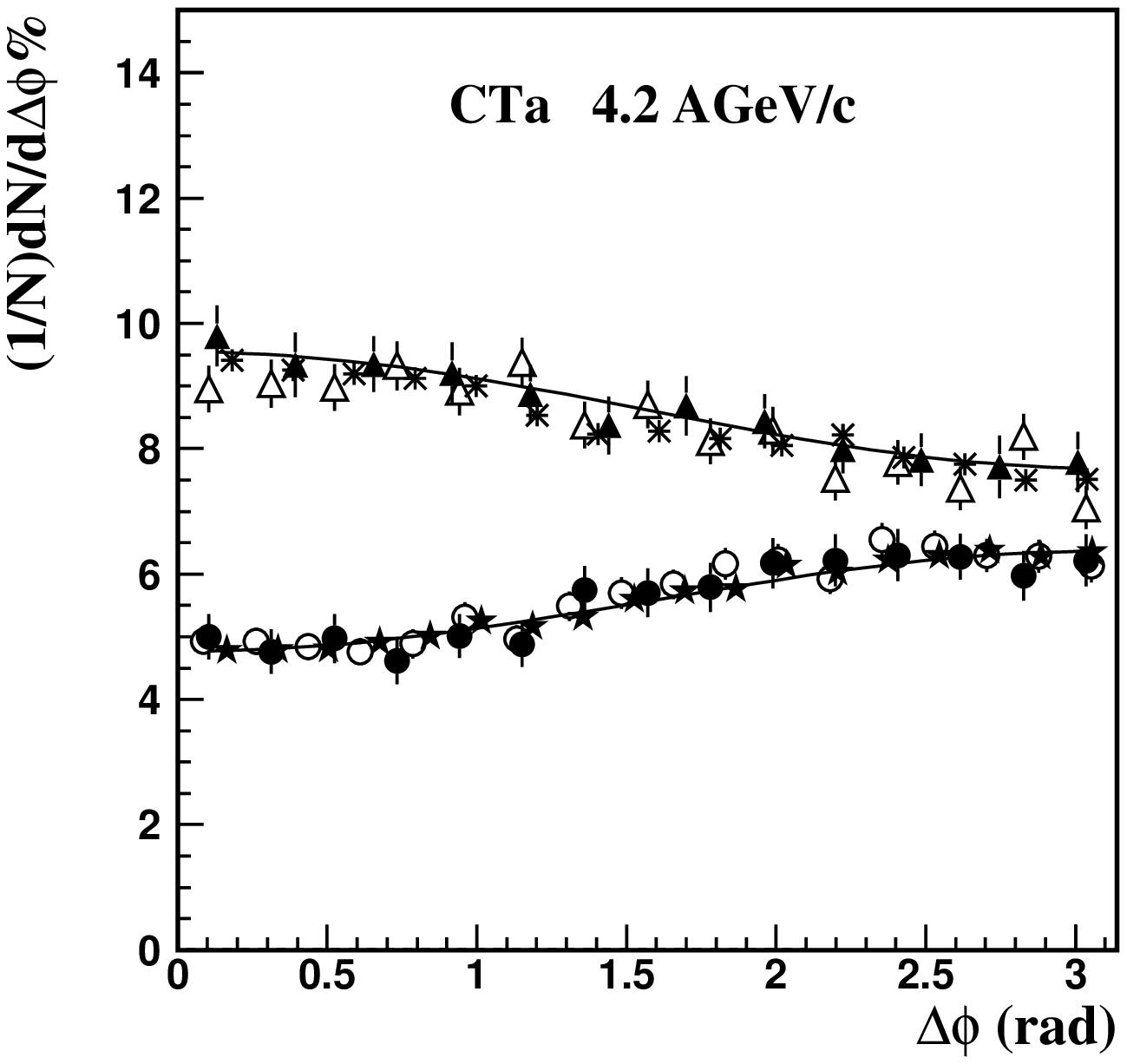}
\caption{}
\end{center}
\end{figure}

\vspace{3cm}

%FIGURE 4
\begin{figure}[bth]
\begin{center}
\includegraphics[width=70mm,height=70mm,clip]{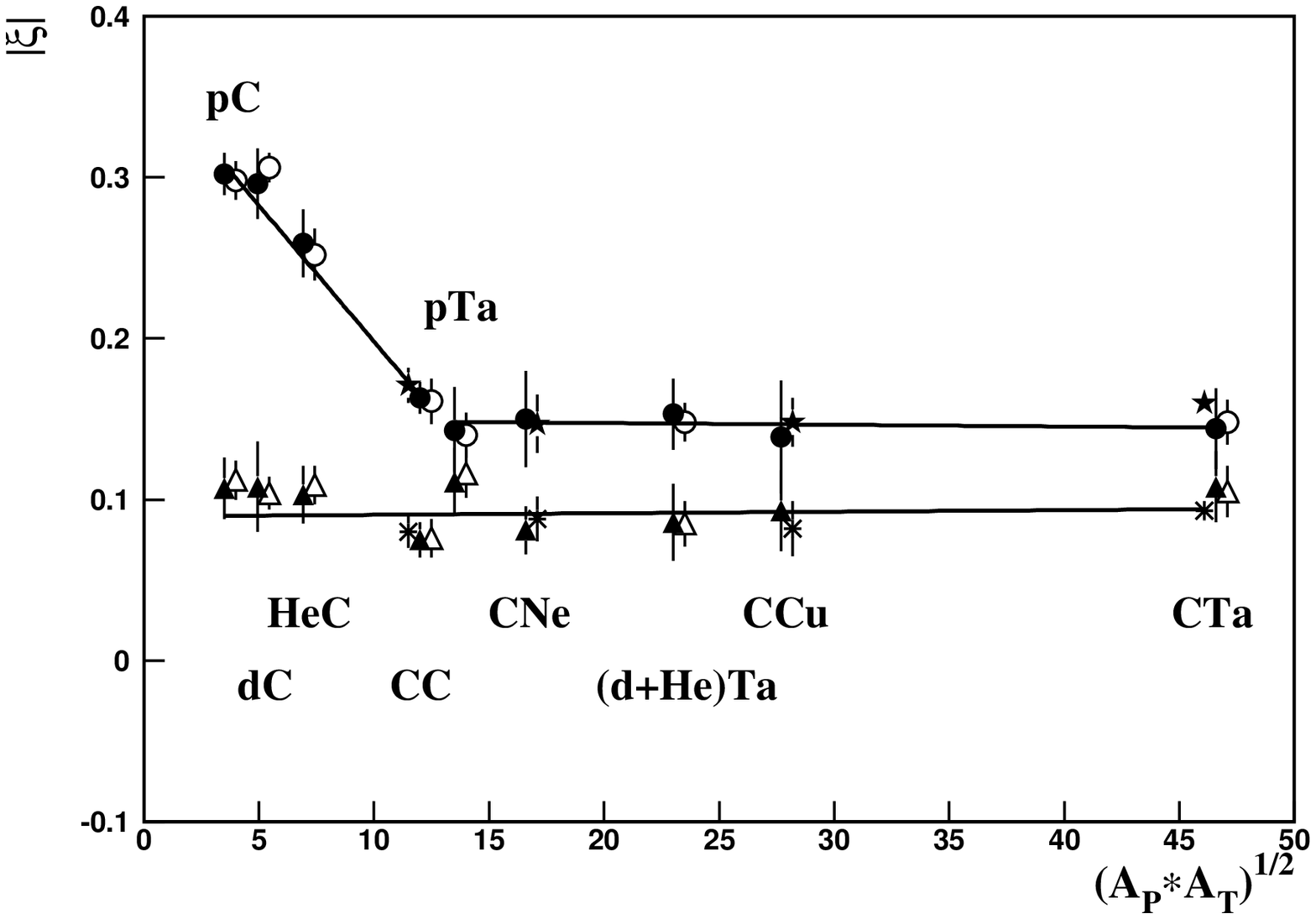}
\caption{}
\end{center}
\end{figure}

\end{document}